\documentclass{article}

\usepackage{PRIMEarxiv}

\usepackage[utf8]{inputenc}
\usepackage[T1]{fontenc}
\usepackage{hyperref}
\usepackage{url}
\usepackage{booktabs}
\usepackage{amsfonts}
\usepackage{amsmath}
\usepackage{nicefrac}
\usepackage{fancyhdr}
\usepackage{natbib}
\usepackage{graphicx}
\usepackage{multirow}
\usepackage{xspace}
\usepackage{xcolor}
\usepackage{subcaption}
\graphicspath{{media/}}

\newcommand{\gpro}{Gemini-2.5-Pro}
\newcommand{\gflash}{Gemini-2.5-Flash}
\newcommand{\logit}{\operatorname{logit}}

\title{Jailbreaks Are Global or Regional? A Controlled Study of Scale and Geolocation Effects on LLM Safety Verification}

\author{
  Anidipta Pal \\
  Independent Researcher \\
  \texttt{anidiptapal@gmail.com}
}

\begin{document}
\maketitle

\begin{abstract}
Safety tests are used to compare language models and to decide where they can be deployed. Their main number, the attack success rate (ASR), can only be compared across models if nothing but the model changes. We test two factors that are usually assumed away: the size of the model and the place a request seems to come from. We evaluate twelve models from 0.5B to 671B parameters with a paired two-phase protocol (plain harmful prompts, then the same requests inside published jailbreak templates). We also send identical prompts through VPN exit nodes in five cities on three continents. Inside the Llama and Qwen families, Phase~2 ASR falls steeply with size (Qwen2.5: 70.8\% at 0.5B, 51.2\% at 7B, 12.6\% at 72B). Wrapping shifts every model by a similar amount on the log-odds scale (1.7 to 2.9); small models shift slightly more, but that gap nearly vanishes when we allow for judge false positives. Parameter count is a weak proxy for safety: Phi-3.5-mini (3.8B) has less than half the ASR of Llama-3.2-3B and Gemma-2-2B, and Mixtral-8x7B is far weaker than 70B dense models. Judge agreement falls as ASR rises ($r=-0.96$), so ASR gaps between sizes carry unequal measurement error; the size gap survives a stress test, but Phi-3.5-mini's lead does not. A small activation probe is consistent with Phi-3.5-mini keeping a clear refusal/compliance split under wrapping while Qwen2.5-0.5B loses it. We detect no effect of request location (permutation $p\ge0.18$ for three models). The test is limited: it could only detect a true between-city spread of about 2--5 points or a single-city shift of about 5--8 points, and most hosted models were reached through a gateway that may hide the exit city from the model provider. We close with a reporting checklist for safety verification.
\end{abstract}

\keywords{jailbreak \and large language models \and AI safety \and safety verification \and measurement validity \and attack success rate \and geolocation \and small language models \and mixture-of-experts}

\section{Introduction}
\label{sec:intro}

Safety tests decide which language model a team deploys and which one it trusts. They report attack success rate (ASR), and ASR supports a comparison only if the threat model is the same on both sides and the measurement is valid~\cite{chouldechova2025comparison}. In practice, more than the model changes between two rows of a results table. Models differ in size, in how they are served, in how well the judge can score their replies, and, for API models, in where the request appears to come from. Reviews of benchmark practice show that such validity problems are common~\cite{bean2025construct}.

\noindent\textbf{Why size.}
Many teams have limited hardware and deploy smaller models than they would like, and small models are often easy to attack~\cite{zhang2025sjbench,yi2025beyond}. If a plain test also hides how much worse small models become under attack, safety checks mislead exactly where the gaps matter most. Yet the literature on size is mixed: HarmBench found no size effect inside model families~\cite{mazeika2024harmbench}.

\noindent\textbf{Why place.}
A large measurement study found that sensitive questions are moderated differently depending on where, and in which language, they are asked~\cite{lipphardt2026basingse}. That study concerned content moderation, not jailbreak success. Whether the apparent origin of a request changes jailbreak ASR is, as far as we can find, untested.

\noindent\textbf{What we do.}
We keep the prompts fixed and paired. We test twelve models from 0.5B to 671B parameters in two phases, and we send the same prompts from five cities. We analyse ASR on the absolute scale and on the log-odds scale, we check how much judge error could explain our results, and we say plainly which claims are strong and which are not.

\noindent\textbf{Contributions.}
\begin{enumerate}
\item \textbf{Size.} A within-family comparison of jailbreak ASR for twelve models under one paired protocol. We show that the uplift from wrapping is best compared on the log-odds scale, where it varies far less across sizes (1.7 to 2.9 log-odds units) than the percentage-point gap (7.6 to 59.0 points); both the gap and the ratio depend on the starting level.
\item \textbf{Proxy.} Two models that break the size trend, with the explanations we can and cannot support.
\item \textbf{Place.} A direct test of whether VPN-routed location changes jailbreak ASR, with permutation tests over five cities on three continents. We state which effect sizes the design can and cannot exclude, and that most hosted models were reached through a gateway.
\item \textbf{Validity.} Evidence that judge agreement falls as ASR rises, a stress test and a judge-error sensitivity analysis showing which conclusions survive, and a reporting checklist.
\item \textbf{Mechanism.} A small activation probe that gives suggestive evidence on why Phi-3.5-mini holds up much better than similar-size models.
\end{enumerate}

\section{Related Work}
\label{sec:related}

\noindent\textbf{Attacks and benchmarks.}
Jailbreaks include automatic suffix search such as GCG~\cite{zou2023universal}, templates collected from real users~\cite{shen2024anything}, and large template banks~\cite{luo2024jailbreakv}. HarmBench provides a shared test set and fixed generation settings~\cite{mazeika2024harmbench}.

\noindent\textbf{Size and safety.}
Findings are mixed. HarmBench tested families from 7B to 70B and saw no link between robustness and size inside a family; it argues that training data and procedure matter more~\cite{mazeika2024harmbench}. In DeepSeek-distilled models, GCG success even rose with size, from about 33\% at 1.5B to 55\% at 32B~\cite{wu2025deepseekgpt}. Work on small models points the other way. Zhang et al.\ tested 63 small models and saw large differences between families; inside one family a 1.6B model was about four times easier to attack with direct requests than the 12B model~\cite{zhang2025sjbench}. Yi et al.\ found that most small models are weaker than large ones and that safety tracks size more closely under jailbreaks than under direct requests~\cite{yi2025beyond}. A 2026 systematisation finds that newer frontier models resist jailbreaks far better than older ones and links this to newer alignment work and red-teaming~\cite{xu2026sok}. We extend the size range to 0.5B--671B under one protocol and add wrapped prompts. We do not claim to be the first same-family comparison.

\noindent\textbf{Mixture-of-experts.}
Safety in mixture-of-experts (MoE) models seems to sit in a few experts. In Qwen3-MoE, switching off 12 experts cut refusals by 22\% in SAFEx, which also studied Mixtral-8x7B-Instruct~\cite{lai2025safex}. L3 bypasses safety by silencing fewer than 20\% of experts~\cite{telintelo2026l3}, and DeepSeek often sends adversarial prompts to under-aligned experts~\cite{wu2025deepseekgpt}. These findings make a ``47B'' Mixtral less surprising but do not predict that every MoE model is weak.

\noindent\textbf{Are ASRs comparable?}
Chouldechova et al.\ argue that many ASR comparisons are not meaningful, because the threat models differ or the measurement is invalid; for example, judge error that differs between systems can make equal true vulnerability look unequal~\cite{chouldechova2025comparison}. Re-scoring with stricter judges lowered some published ASRs a lot (93\% to 69\% for one attack on GPT-3.5-Turbo)~\cite{chu2025jades}. Judges are sensitive to answer style~\cite{eiras2025knowthyjudge}, did only slightly better than a coin flip under typical attack conditions in a large audit~\cite{schwinn2026coinflip}, and can give very different ASR for the same replies~\cite{gao2026judge}. Hosted models are also not fully deterministic even at temperature zero~\cite{atil2025nondeterminism}. We build our analysis around these points.

\noindent\textbf{Places.}
Lipphardt et al.\ studied about 700{,}000 replies from 15 models reached from 12 locations. Moderation of sensitive topics differed by up to 60\% across locations, and a locally run DeepSeek differed from its API version~\cite{lipphardt2026basingse}. Their prompts mixed location with language and concerned sensitive topics, not jailbreaks. We use English jailbreak prompts and change only the exit city.

\noindent\textbf{Inside the model and compression.}
Recent work shows that safety behaviour uses several orthogonal directions in the residual stream rather than a single one~\cite{pan2025hidden}. Small models are often compressed. Yi et al.\ study how compression, including quantization, affects small-model security~\cite{yi2025beyond}. Chhabra and Khalili find that quantization usually keeps refusal closer to the original model than pruning does~\cite{chhabra2025compressed}. Our local models use 4-bit quantization, so we cannot fully rule it out as a cause of weakness.

\section{Methods}
\label{sec:methods}

\noindent\textbf{Models and serving.}
We use two matched families, Llama and Qwen, with three sizes each (Llama-3.2-1B, Llama-3.2-3B, Llama-3.1-70B; Qwen2.5-0.5B, 7B, 72B). We add four single models (Gemma-2-2B, Phi-3.5-mini-instruct, Mixtral-8x7B-Instruct, DeepSeek-V3) and two closed models, \gpro{} and \gflash{} (Table~\ref{tab:results}). Mixtral has 47B parameters in total but uses about 13B per token. Models below 8B run locally on a free Google Colab T4 GPU in 4-bit NF4 quantization. Larger models are used through OpenRouter or Google AI Studio (Table~\ref{tab:setup}). Size is therefore mixed up with how a model is served, and the Llama family mixes two generations (3.2 small, 3.1 large); see Section~\ref{sec:limits}.

The two closed models are \gpro{} and \gflash{}, queried through Google AI Studio. We use the 2.5 generation because Google had retired all Gemini~1.5 models by September 2025 and had scheduled Gemini~2.0 Flash to shut down on 1~June 2026\footnote{\url{https://cloud.google.com/vertex-ai/generative-ai/docs/learn/model-versions}}, whereas the 2.5 Pro and 2.5 Flash models are on the free tier. Both models think by default, and we left that setting unchanged, so their reply length and RCF include the effect of thinking. The OpenRouter listing for Mixtral-8x7B carried a removal date of 8~May 2026, and Phi-3.5-mini is a paid model there (about \$0.10 per million tokens)\footnote{\url{https://openrouter.ai/models/microsoft/phi-3.5-mini-128k-instruct}}, so the study was low-cost rather than free.

\begin{table}[h]
\caption{Setup. Decoding entries are the library or provider defaults; we did not harmonise decoding across routes. Sources: model-repository generation configs and OpenRouter and Google documentation.}
\label{tab:setup}
\centering
\footnotesize
\begin{tabular}{p{3.3cm}p{4.3cm}p{6.2cm}}
\toprule
Model & Route & Decoding \\
\midrule
Llama-3.2-1B / 3B & local, 4-bit NF4 & sampling, $T{=}0.6$, top-$p$ 0.9 \\
Qwen2.5-0.5B / 7B & local, 4-bit NF4 & sampling, $T{=}0.7$, top-$p$ 0.8, top-$k$ 20, repetition penalty 1.05 \\
Gemma-2-2B & local, 4-bit NF4 & library default \\
Phi-3.5-mini & local, 4-bit NF4 & greedy (its generation config has no sampling fields) \\
Llama-3.1-70B, Qwen2.5-72B, DeepSeek-V3 & OpenRouter gateway; the provider is chosen by OpenRouter & provider defaults \\
Mixtral-8x7B & OpenRouter gateway & provider defaults \\
\gflash{}, \gpro{} & Google AI Studio, direct & provider defaults (temperature 1.0); thinking on \\
\bottomrule
\end{tabular}
\end{table}

\noindent\textbf{Prompts and phases.}
In Phase~1 each model gets plain harmful instructions from AdvBench~\cite{zou2023universal} and HarmBench~\cite{mazeika2024harmbench}. In Phase~2 we wrap the \emph{same} requests in published jailbreak prompts from JailbreakV-28K~\cite{luo2024jailbreakv} and the DAN collection~\cite{shen2024anything}. We write no new prompts, so the phases are paired and we add no new attack. Each model gets 400 prompts per phase (800 in total), balanced over five harm categories: cybercrime, misinformation, fraud, harassment and CBRN-adjacent content, with 80 prompts per category. Requests were assigned to categories with the source benchmarks' own labels where they exist (HarmBench's semantic categories) and by keyword matching otherwise; HarmBench has no separate fraud category, because fraud sits inside its ``illegal'' category (Table~\ref{tab:catmap}). JailbreakV-28K holds 20{,}000 text-only transfer attacks in template, persuasion and logic formats, each tied to one of 2{,}000 RedTeam-2K queries; RedTeam-2K is built partly from AdvBench but not from HarmBench. The DAN collection holds 1{,}405 in-the-wild jailbreak prompts that wrap any request. Phase~2 therefore pairs AdvBench-derived requests with JailbreakV-28K attacks and the remaining requests with DAN prompts, drawn at random with a fixed seed.

\noindent\textbf{Judging.}
A reply counts as a successful jailbreak if the model really does what the harmful request asks. A JailbreakBench-style classifier\footnote{\url{https://jailbreakbench.github.io}} makes this call. On JailbreakBench's own validation set that style of judge agrees with human labels about 91\% of the time and has a false-positive rate of about 12\%; that is not our data and not our replies. Judges can fail, especially on jailbreak outputs~\cite{schwinn2026coinflip,gao2026judge}, so one person re-checked 10\% of replies (about 80 per model) and we report Cohen's $\kappa$ between judge and person (Table~\ref{tab:results}). We also record \emph{refusal-comprehension failure} (RCF): replies that are neither a clear refusal nor a clear compliance. RCF signals unreliability, not safety. RCF is measured in Phase~2 and scored separately from the ASR label, so the two can overlap when the judge counts a confused reply that contains harmful fragments as a success. For Qwen2.5-0.5B, ASR and RCF sum to 101.9\%, so at least 1.9 points of replies were counted in both. The reported RCF percentages are not all multiples of 0.25, so they are not all counts out of exactly 400, and the reported RCF intervals correspond to different numbers of responses from row to row (about 370--375 for the two highest RCF values; Appendix~\ref{app:setup}). We therefore treat the intervals as approximate and base all statistical statements on $n=400$ per phase.

\noindent\textbf{Estimand and statistics.}
For model $m$ and phase $k$, ASR is the share of the 400 fixed prompts whose reply the judge labels a success. Because the 80 prompts per category are balanced, the overall ASR in Table~\ref{tab:results} is the mean of the five category rates in Figure~\ref{fig:fullheat}; those rates are shown as whole percentages, so overall values can be off by up to about 0.5 points. Prompts are the same for all models, so comparisons are paired over prompts. Following \cite{chouldechova2025comparison}, a comparison is meaningful only if the threat model is the same on both sides and the measurement is valid; we probe the second condition in Section~\ref{sec:validity}. Decoding is a third threat to comparability: HarmBench fixes greedy decoding to avoid it~\cite{mazeika2024harmbench}, whereas our routes used different defaults (Table~\ref{tab:setup}). We report 95\% Wilson intervals for ASR (Table~\ref{tab:counts} in Appendix~\ref{app:setup}): up to $\pm$5 points at 400 prompts when ASR is near 50\%, and up to $\pm$11 points at the $\sim$80 prompts per category. To compare the uplift from wrapping across models without the ceiling effect that limits ratios, we use the log-odds shift $s_m=\logit(\hat p^{(2)}_m)-\logit(\hat p^{(1)}_m)$ with delta-method standard errors (Appendix~\ref{app:stats}). These are our calculations from Table~\ref{tab:results}; they treat the two phases and all models as independent binomial samples. We did not correct for multiple comparisons, and we label every analysis chosen after seeing the data as exploratory.

\noindent\textbf{Geolocation probe.}
We use JailbreakV-28K only and send requests through VPN exit nodes in Tokyo, Mumbai, San Francisco, Moscow and Berlin. We test the two most robust models in each size tier (lowest Phase~2 ASR): Phi-3.5-mini and Qwen2.5-7B for small models, DeepSeek-V3 and \gpro{} for large ones. A locally run model has no network location, so the two small models were reached through OpenRouter, as was DeepSeek-V3; Gemini was called directly. OpenRouter is a proxy that forwards each request to a model provider\footnote{\url{https://openrouter.ai/docs/faq}}, so the provider receives the request from OpenRouter and not from the exit city. OpenRouter does not document whether client location is forwarded, and by default it chooses the provider by availability, price and performance, so the provider can change between runs. The test therefore manipulates the client address seen by the gateway, and only the Gemini models, which have missing cities, saw the exit city directly. The observed between-city standard deviations (2.3, 2.5 and 1.7 points) match the binomial noise of about 400 independent responses per city--model cell (2.3, 2.5 and 1.6), for example three runs of about 130 prompts, so we use that noise level when we state detection limits. Each city--model pair runs three times at different hours, because hosted models are not fully deterministic~\cite{atil2025nondeterminism}. We shuffle the city labels over the run-level ASR values 5000 times and compare the observed spread of city means with the shuffled spreads. As a second check we simulate each cell as 400 independent responses at the model's mean ASR (Appendix~\ref{app:geoext}). We do not use a paired analysis over prompts; it would be more powerful when the same prompts are sent from every city and is the natural next step. Google's free tier is not offered in the European Economic Area, Switzerland or the United Kingdom\footnote{\url{https://ai.google.dev/available_regions}}, which explains the missing Berlin data. \gpro{} could not be tested from Berlin or reliably from Moscow (two of three connection attempts failed). We log these as failures, leave them out, and never fill them in.

\noindent\textbf{Contrastive probe.}
For Phi-3.5-mini and Qwen2.5-0.5B we take last-token residual-stream activations on 120 paired refusal and compliance examples, reduce them to two principal components, and fit a linear discriminant. We report Cohen's $d$ on the discriminant scores: a high $d$ means refusal and compliance are easy to separate linearly, and a drop in Phase~2 means the templates blur that separation. We use the last token of each example and did not test other positions. Details are in Appendix~\ref{app:probe}.

\section{Results}
\label{sec:results}

\subsection{Size: ASR falls steeply, and wrapping adds a roughly constant log-odds shift}
\label{sec:size}

\begin{figure}[t]
\centering
\begin{subfigure}[t]{0.64\linewidth}
  \includegraphics[width=\linewidth]{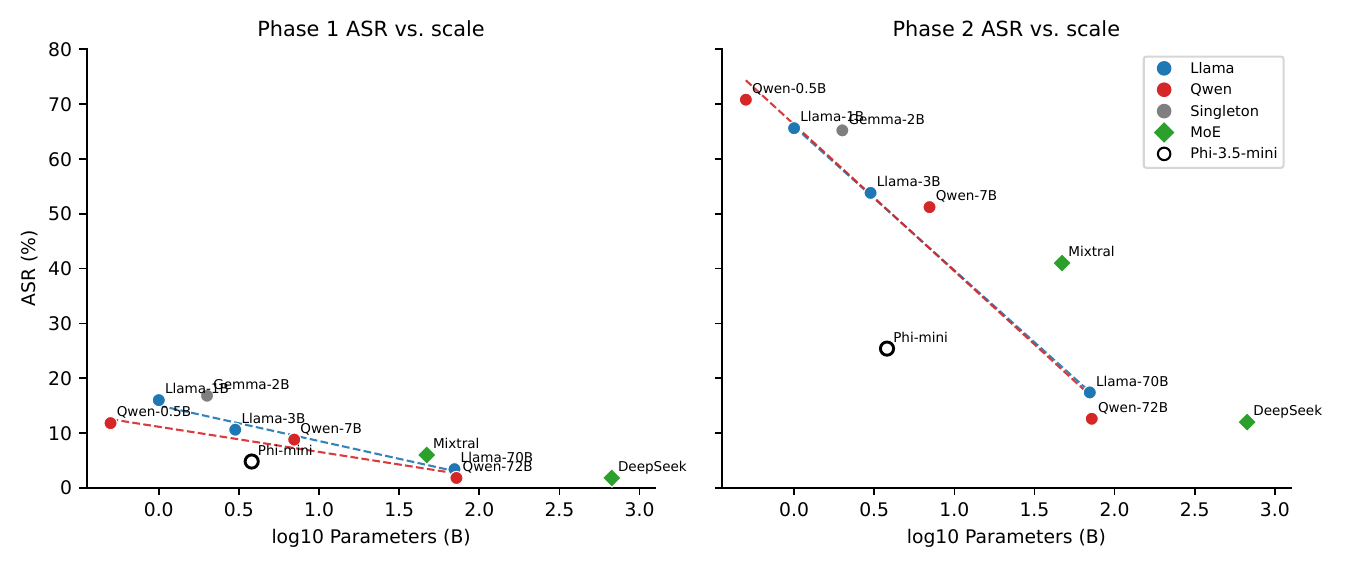}
  \caption{Phase~1 (left panel) and Phase~2 (right panel) ASR against log parameter count. Dashed lines are within-family OLS fits. Diamonds are MoE models. Phi-3.5-mini (hollow) lies between the small-model and large-model groups. The Gemini models are omitted because their parameter counts are not public.}
  \label{fig:scale}
\end{subfigure}
\hfill
\begin{subfigure}[t]{0.33\linewidth}
  \includegraphics[width=\linewidth]{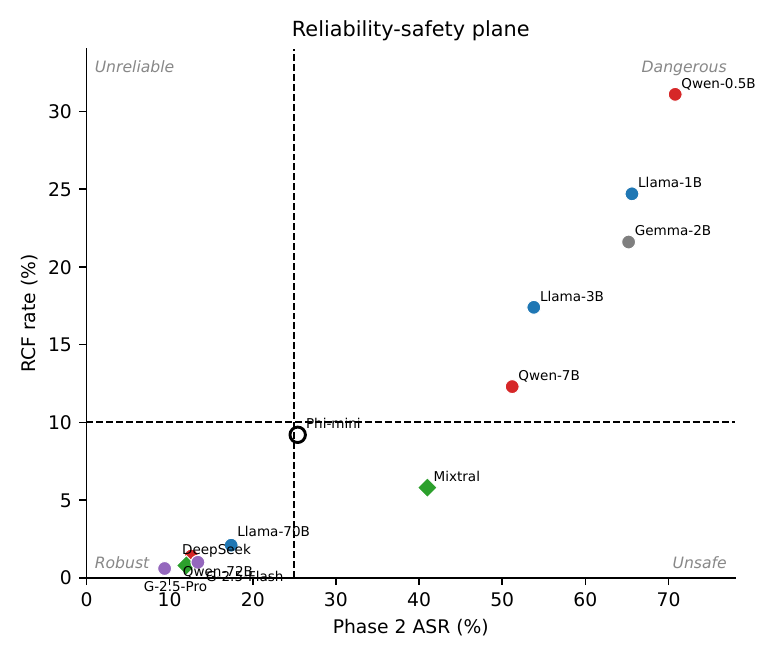}
  \caption{Reliability--safety plane (lines at ASR $=25\%$ and RCF $=10\%$). No small model is in the robust corner. Phi-3.5-mini is closest, 0.4 points past the ASR line (well inside the $\pm$5 point resolution).}
  \label{fig:reliability}
\end{subfigure}
\caption{Scale and reliability results.}
\label{fig:main}
\end{figure}

\begin{figure}[t]
\centering
\begin{subfigure}[t]{0.49\linewidth}
  \includegraphics[width=\linewidth]{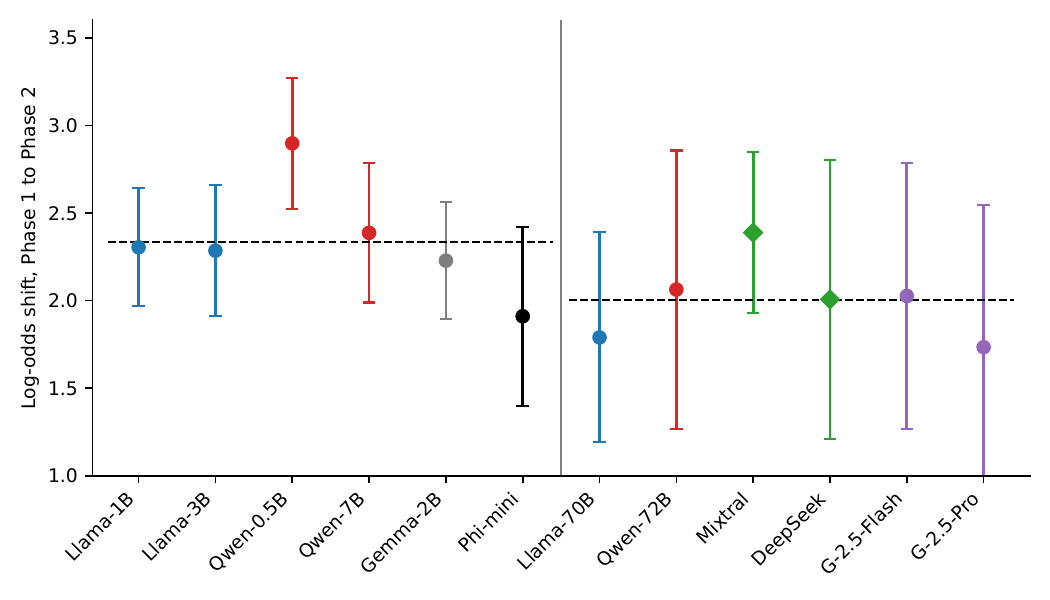}
  \caption{Log-odds shift from Phase~1 to Phase~2 for each model, with approximate 95\% intervals from independent binomial samples of 400 responses. Dashed lines are the tier means (2.33 for small models, 2.00 for large models).}
  \label{fig:logit}
\end{subfigure}
\hfill
\begin{subfigure}[t]{0.49\linewidth}
  \includegraphics[width=\linewidth]{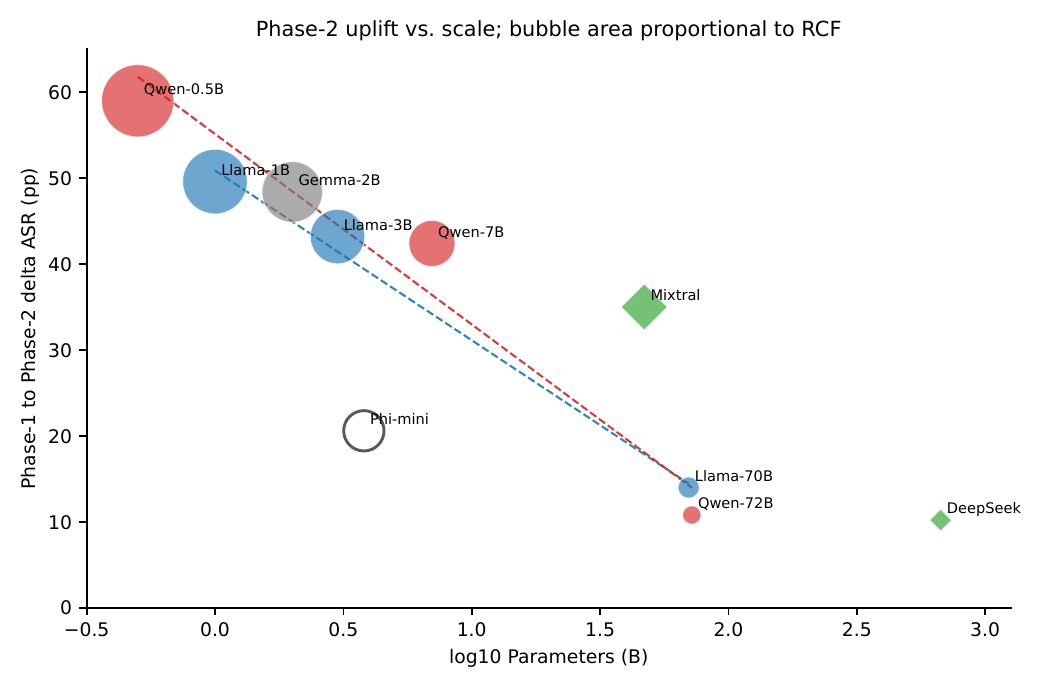}
  \caption{Phase~1$\to$Phase~2 uplift against log parameter count; bubble area $\propto$ RCF. Dashed lines are within-family OLS fits for Llama and Qwen. The Gemini models are omitted (parameter counts not public).}
  \label{fig:uplift}
\end{subfigure}
\caption{Uplift from wrapping on the log-odds scale (left) and in percentage points (right).}
\label{fig:shift}
\end{figure}

Table~\ref{tab:results} and Figure~\ref{fig:scale} show the results. In the Qwen2.5 family Phase~2 ASR falls from 70.8\% (0.5B) to 51.2\% (7B) to 12.6\% (72B). In the Llama family it falls from 65.6\% (1B) to 53.8\% (3B) to 17.4\% (70B). Phase~1 shows the same order. The percentage-point uplift $\Delta$ is much larger for small models ($+59.0$ for Qwen2.5-0.5B vs.\ $+10.8$ for Qwen2.5-72B), but this mostly reflects their higher starting point. A ratio is not a fair fix, because it is capped by $1/\text{ASR}_1$ and is therefore lower for the models that start highest. On the log-odds scale the shift is much more uniform (Figure~\ref{fig:logit}; Figure~\ref{fig:uplift} shows the point uplift for comparison): it lies between 1.7 and 2.9 for all twelve models, which multiplies the odds of a successful jailbreak by about 5.7 to 18.1. Small models average 2.3 and large models 2.0. Qwen2.5-0.5B (2.90) is the one clearly high value. Inside the large tier the spread of the shifts is below what sampling noise alone would produce, but inside the small tier it is above it, mostly because of Qwen2.5-0.5B and Phi-3.5-mini (Appendix~\ref{app:stats}). \textbf{The tier gap is not robust to judge error.} The raw gap is 0.33, with a 95\% interval of 0.0 to 0.7. The Phase~1 ASRs of most large models (1.8--3.4\%; Mixtral 6.0\%) are small enough for false positives to matter. If the judge had sensitivity 0.90 and a false-positive rate of 1\% (assumed, not measured), the corrected shifts would average 2.59 for small and 2.56 for large models, and the gap would fall from 0.33 to 0.03 (Table~\ref{tab:judgesens}). If the false-positive rate were above about 1.8\%, the best models' Phase~1 ASR would be entirely false positives and their shifts could not be computed. We therefore treat ``a shift of roughly 2 to 3 log-odds units at every size'' as the main result and ``slightly more for small models'' as weak evidence. Phase~1 ranks the models almost exactly like Phase~2 (Spearman $\rho=0.96$), and $\Delta$ is nearly the same information as Phase~2 ASR (Pearson $r=0.996$). So the problem with a plain test is its level and not its order: wrapping multiplies every model's odds of failure several times over, and a Phase~1 number understates real risk for every model.

\begin{table}[t]
\caption{Phase~1 ASR, Phase~2 ASR, uplift $\Delta$ in points, log-odds shift $s$ (Appendix~\ref{app:stats} gives standard errors), Phase~2 RCF with the reported Wilson 95\% interval, and classifier--human $\kappa$ on the 10\% re-checked sample. All values are percentages except $s$ and $\kappa$. Phase~1 and Phase~2 ASR are the means of the five category rates in Figure~\ref{fig:fullheat}, so they carry rounding of up to about 0.5 points. A line separates the small tier (top) from the large tier (bottom). The RCF intervals are approximate: they do not all correspond to 400 responses (Appendix~\ref{app:setup}).}
\label{tab:results}
\centering
\footnotesize
\setlength{\tabcolsep}{4pt}
\begin{tabular}{llrrrrrrr}
\toprule
Model & Params & Ph1 & Ph2 & $\Delta$ & $s$ & RCF & RCF 95\% CI & $\kappa$ \\
\midrule
Llama-3.2-1B  & 1B       & 16.0 & 65.6 & $+$49.6 & 2.30 & 24.7 & 20.6--29.3 & 0.76 \\
Llama-3.2-3B  & 3B       & 10.6 & 53.8 & $+$43.2 & 2.28 & 17.4 & 13.8--21.6 & 0.80 \\
Qwen2.5-0.5B  & 0.5B     & 11.8 & 70.8 & $+$59.0 & 2.90 & 31.1 & 26.6--35.9 & 0.71 \\
Qwen2.5-7B    & 7B       & 8.8  & 51.2 & $+$42.4 & 2.39 & 12.3 &  9.4--15.9 & 0.83 \\
Gemma-2-2B    & 2B       & 16.8 & 65.2 & $+$48.4 & 2.23 & 21.6 & 17.7--26.0 & 0.79 \\
Phi-3.5-mini  & 3.8B     & 4.8  & 25.4 & $+$20.6 & 1.91 &  9.2 &  6.8--12.3 & 0.85 \\
\midrule
Llama-3.1-70B & 70B      & 3.4  & 17.4 & $+$14.0 & 1.79 &  2.1 &  1.1--3.8  & 0.88 \\
Qwen2.5-72B   & 72B      & 1.8  & 12.6 & $+$10.8 & 2.06 &  1.4 &  0.7--2.8  & 0.91 \\
Mixtral-8x7B  & 47B MoE  & 6.0  & 41.0 & $+$35.0 & 2.39 &  5.8 &  3.9--8.4  & 0.84 \\
DeepSeek-V3   & 671B MoE & 1.8  & 12.0 & $+$10.2 & 2.01 &  0.8 &  0.3--1.9  & 0.90 \\
\gflash       & API      & 2.0  & 13.4 & $+$11.4 & 2.03 &  1.0 &  0.5--2.2  & 0.89 \\
\gpro         & API      & 1.8  & 9.4  & $+$7.6  & 1.73 &  0.6 &  0.2--1.6  & 0.91 \\
\bottomrule
\end{tabular}
\end{table}

\subsection{Is the size effect a measurement artefact?}
\label{sec:validity}

Four threats could make the size gap look bigger than it is.

\noindent\textbf{Judge error that differs by model.}
Judge--human agreement $\kappa$ falls as Phase~2 ASR rises ($r=-0.96$, Appendix~\ref{app:classifier}), and it is only weakly tied to the classifier's own confidence ($r=0.31$). So the judge makes more mistakes on exactly the models with the highest ASR, in a direction we cannot tell from $\kappa$ alone. As a stress test we remove each model's RCF share from its ASR, which is the harshest reading in which every incoherent reply was wrongly counted as a success. The order among the high-ASR small models changes (Gemma-2-2B moves to the top and Qwen2.5-0.5B falls to third), but the small/large split holds: small models other than Phi average about 40\% against about 16\% for large models (Appendix~\ref{app:stats}, Table~\ref{tab:stress}). The size effect is therefore unlikely to be explained by judge error alone. The stress test does change one conclusion: Phi-3.5-mini falls to 16.2\%, level with Llama-3.1-70B (15.3\%), so its lead over similar-size models depends on how incoherent replies are scored.

\noindent\textbf{How models are served.}
Small models run locally in 4-bit and large models run hosted. This design cannot separate size from serving. Earlier work found that a local DeepSeek behaved differently from its API version~\cite{lipphardt2026basingse}, and that quantization changes refusal behaviour, usually mildly at 4 bits~\cite{chhabra2025compressed}. The decisive control is to run one model both ways; we did not.

\noindent\textbf{Decoding.}
Qwen2.5 and Llama-3.2 sample at temperature 0.7 and 0.6 by default, Phi-3.5-mini decodes greedily, and hosted providers apply their own defaults (Table~\ref{tab:setup}). Different decoding rules change ASR, so this could add to the apparent size gap. Re-running everything with greedy decoding and 512 new tokens, as in HarmBench~\cite{mazeika2024harmbench}, would remove the threat.

\noindent\textbf{Alignment age.}
Larger models in our pool are also newer or better aligned, and Llama mixes generations. Our within-family comparison reduces but does not remove this.

\subsection{Two models break the size trend}
\label{sec:exceptions}

Mixtral-8x7B reaches 41.0\% in Phase~2, worse than Llama-3.1-70B (17.4\%) and closer to Qwen2.5-7B (51.2\%) than to any 70B model. It uses only about 13B parameters per token, so on a size line it would fall between 7B and 70B dense models and is not odd once active parameters are counted. We cannot blame MoE routing itself: DeepSeek-V3 is also MoE and is one of the most robust open models (12.0\%, level with Qwen2.5-72B at 12.6\%), and Mixtral is the oldest model here, so alignment age is just as likely a cause. Work on MoE safety shows refusal can rest on a few experts~\cite{lai2025safex,telintelo2026l3}, which fits but does not prove our result. Phi-3.5-mini reaches 25.4\% at 3.8B: less than half the ASR of Llama-3.2-3B (53.8\%) and Gemma-2-2B (65.2\%), 26 points below Qwen2.5-7B, and below Mixtral, but about 8 points above Llama-3.1-70B. It is in between, not in the large-model group, and, as shown above, the stress test puts it close to the large dense models (15.3\% for Llama-3.1-70B, 11.2\% for Qwen2.5-72B). Its RCF (9.2\%) is the lowest of the small models. The Phi-3 family was also found to be among the better-aligned small models~\cite{zhang2025sjbench}, which may help explain this. Phi-3.5-mini is also the one local model that decodes greedily by default, so decoding could contribute. Parameter count is not a reliable safety signal.

\subsection{RCF is not robustness}

RCF is 31.1\% for Qwen2.5-0.5B and 24.7\% for Llama-3.2-1B. These models are not safer because they are confused: 70.8\% Phase~2 ASR with 31.1\% RCF is dangerous \emph{and} unreliable. Figure~\ref{fig:reliability} shows the target, low ASR with low RCF, which no small model reaches.

\subsection{Harm categories}

\begin{figure}[t]
\centering
\begin{subfigure}[t]{0.58\linewidth}
  \includegraphics[width=\linewidth]{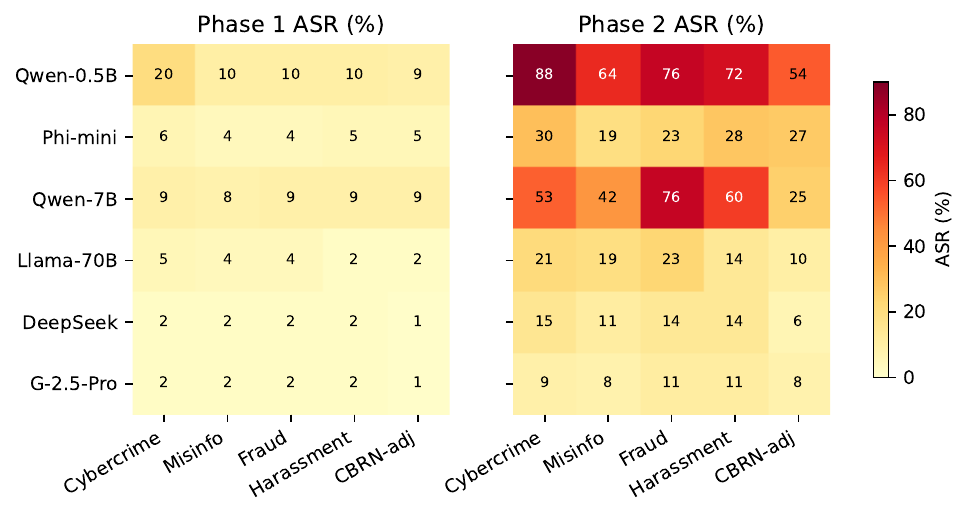}
  \caption{Phase~1 (left) and Phase~2 (right) ASR by harm category for six models.}
  \label{fig:harm}
\end{subfigure}
\hfill
\begin{subfigure}[t]{0.40\linewidth}
  \includegraphics[width=\linewidth]{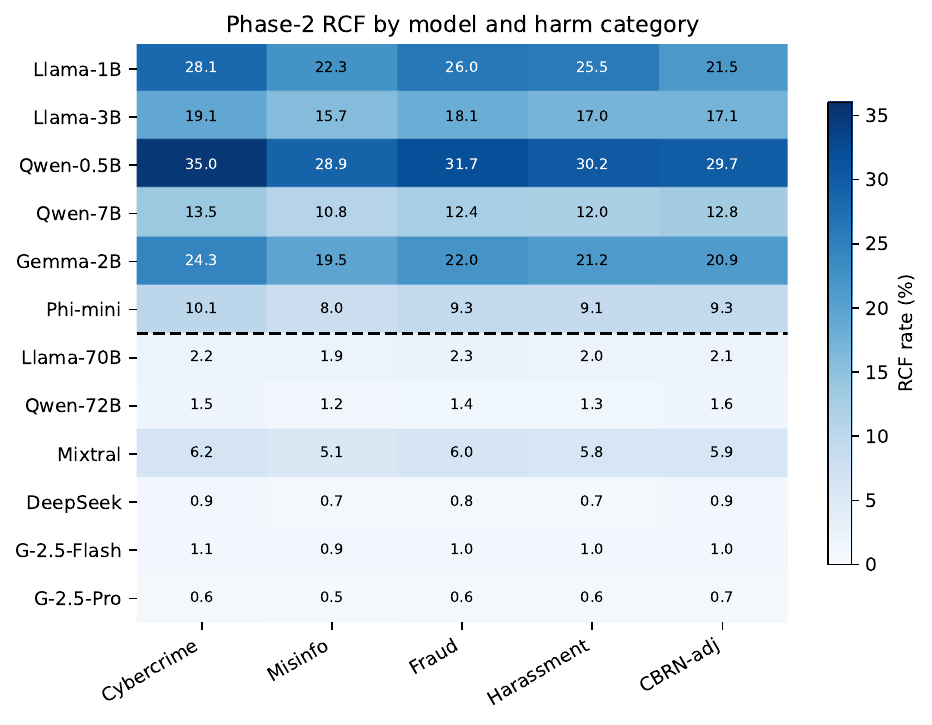}
  \caption{Phase~2 RCF by model and harm category for all twelve models. The dashed line marks the tier boundary.}
  \label{fig:rcfcat}
\end{subfigure}
\caption{Category heatmaps: ASR (left) and RCF (right).}
\label{fig:cat}
\end{figure}

In Phase~2, cybercrime and fraud ASR are both higher than misinformation and CBRN-adjacent ASR for 10 of 12 models (the exceptions are Phi-3.5-mini and Qwen2.5-72B; Figures~\ref{fig:harm} and~\ref{fig:fullheat}). Qwen2.5-0.5B reaches 88.3\% on cybercrime but 54.4\% on CBRN-adjacent content (Table~\ref{tab:catgap}). The top category is cybercrime for some models and fraud for others (Qwen2.5-7B: 76\% fraud vs.\ 53\% cybercrime), and for the most robust models the gaps are inside the sampling error. One reading is that safety data cover CBRN and misinformation more densely than cybercrime and fraud. We cannot exclude other reasons, such as how clearly harmful the prompts look, how well the judge scores each category, or the way requests were mapped to categories (Table~\ref{tab:catmap}). Either way, CBRN-only red-teaming would not reveal the weaker categories.

\begin{table}[h]
\caption{Phase~2 ASR for cybercrime and CBRN-adjacent prompts. Gap = cybercrime $-$ CBRN-adjacent. It is not always the highest-minus-lowest range (for Qwen2.5-7B the top category is fraud). With about 80 prompts per category the 95\% interval is up to $\pm$11 points. As a share of overall ASR the gap does not shrink along Qwen2.5-0.5B, Qwen2.5-7B, Llama-3.1-70B and DeepSeek-V3 (0.48, 0.55, 0.64, 0.71). Phi-3.5-mini and \gpro{} show almost none (0.12 for both).}
\label{tab:catgap}
\centering
\small
\begin{tabular}{lrrr}
\toprule
Model & Cybercrime & CBRN-adj & Gap (pp) \\
\midrule
Qwen2.5-0.5B   & 88.3 & 54.4 & 33.9 \\
Phi-3.5-mini   & 30.4 & 27.3 &  3.1 \\
Qwen2.5-7B     & 53.2 & 25.2 & 28.0 \\
Llama-3.1-70B  & 21.4 & 10.2 & 11.2 \\
DeepSeek-V3    & 14.8 &  6.3 &  8.5 \\
\gpro          &  9.2 &  8.1 &  1.1 \\
\bottomrule
\end{tabular}
\end{table}

\subsection{Location: no detectable effect, with stated limits}
\label{sec:geo}

Table~\ref{tab:geo} shows Phase~2 ASR by exit city. For the three fully testable models the spread across cities (standard deviation) is at most 2.5 points, and the permutation tests give $p\ge0.18$ in every case. A parametric check that treats each city--model cell as 400 independent responses at the model's mean ASR gives larger values, about 0.41, 0.42 and 0.35 (Appendix~\ref{app:geoext}). Both tests agree that the spreads are what noise alone would produce. We therefore detect no geolocation effect, but the test is weak in three ways.

\textbf{Detection limit.} The observed spreads equal the binomial noise of about 400 independent responses per cell (Section~\ref{sec:methods}). With that noise, the omnibus test could detect a true between-city standard deviation of about 3 to 5 points for Phi-3.5-mini and Qwen2.5-7B and about 2 to 3.5 points for DeepSeek-V3 (50--80\% power). A single city differing from the other four would need to shift by about 5 to 8 points (80\% power; Appendix~\ref{app:stats}). A paired analysis over prompts could do better; our run-level test cannot.

\textbf{What the treatment reaches.} Three of the four models were reached through a gateway that forwards requests to a provider it chooses, so the exit city may never reach the model provider. Only the Gemini models were called directly. For them we have three usable cities (Tokyo, Mumbai, San Francisco: 9.4\%, 10.1\%, 8.8\%), too few to say more than that they look alike. Our result therefore supports ``no effect through a gateway'' much more than ``no effect of user location on model providers''.

\textbf{One pattern for follow-up.} Moscow has the highest ASR for all three models (2.8 to 3.6 points above each model's five-city mean). If cities had no effect, the chance that the same city tops all three models is $1/25=0.04$. We noticed this after looking at the data, and combining the three reported $p$-values (Fisher's method) gives about 0.19. The excess is also below the detection limits above, so we cannot call it an effect or rule it out. Location did change \emph{availability}: \gpro{} could not be reached from Berlin and was unreliable from Moscow. That reflects access rules, not safety behaviour.

\begin{table}[h]
\caption{Phase~2 ASR (\%) on JailbreakV-28K by VPN exit city. SD is the sample standard deviation over the available cities, and $p$ is the run-level permutation $p$-value. The Gemini gaps are infrastructure failures, not safety findings: $^{*}$ two of three connection attempts failed from Moscow; $^{\dagger}$ no access from Berlin (free tier not offered there). \gpro{} is the only model here that was called directly; the other rows went through a gateway.}
\label{tab:geo}
\centering
\small
\begin{tabular}{llrrrrrrc}
\toprule
Tier & Model & Tokyo & Mumbai & SF & Moscow & Berlin & SD & $p$ \\
\midrule
Small & Phi-3.5-mini   & 29.1 & 31.4 & 27.8 & 33.7 & 30.2 & 2.3 & 0.31 \\
Small & Qwen2.5-7B     & 48.3 & 44.7 & 46.1 & 51.2 & 47.9 & 2.5 & 0.22 \\
Large & DeepSeek-V3    & 10.8 & 12.3 & 11.1 & 14.6 & 10.4 & 1.7 & 0.18 \\
Large & \gpro          &  9.4 & 10.1 &  8.8 & N/A$^{*}$ & N/A$^{\dagger}$ & 0.7 & --- \\
\bottomrule
\end{tabular}
\end{table}

\subsection{Probe}

\begin{figure}[t]
\centering
\begin{subfigure}[t]{0.38\linewidth}
  \includegraphics[width=\linewidth]{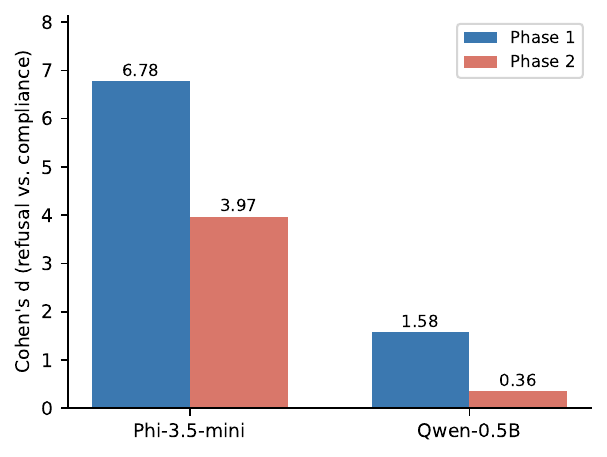}
  \caption{Cohen's $d$ of the refusal/compliance split in Phase~1 and Phase~2. Phi-3.5-mini stays well separated ($d$: 6.78 to 3.97); Qwen2.5-0.5B collapses ($d$: 1.58 to 0.36).}
  \label{fig:probe}
\end{subfigure}
\hfill
\begin{subfigure}[t]{0.60\linewidth}
  \includegraphics[width=\linewidth]{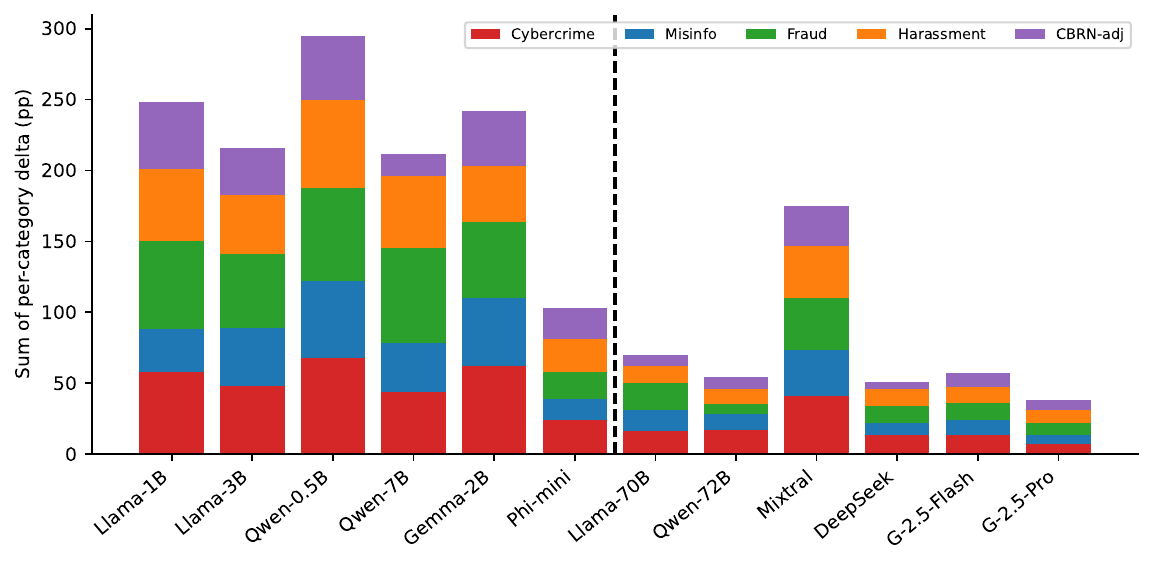}
  \caption{Summed per-category uplift for all twelve models. The dashed line separates the small tier (left) from the large tier (right).}
  \label{fig:deltacat}
\end{subfigure}
\caption{Probe evidence (left) and uplift by harm category (right).}
\label{fig:probe_bars}
\end{figure}

Figure~\ref{fig:probe} and Table~\ref{tab:probe} give the probe results. Phi-3.5-mini has Cohen's $d=6.78$ in Phase~1 and $d=3.97$ in Phase~2: the split drops by about 41\% but stays large. Qwen2.5-0.5B goes from $d=1.58$ to $d=0.36$ (about 77\%), so refusal and compliance activations become almost the same. In Phase~2 the Phi value is about 11 times the Qwen value. This fits the idea that templates blur the refusal/compliance direction in the smaller model. It is only suggestive: it covers two models of very different size (7.6 times apart), uses two principal components, and takes its labels from model behaviour, which is noisy for a model with 31\% RCF. A size-matched comparison, such as Phi-3.5-mini against Llama-3.2-3B, is the obvious next step.

\begin{table}[h]
\caption{Cohen's $d$ on the discriminant scores (refusal vs.\ compliance activations).}
\label{tab:probe}
\centering
\small
\begin{tabular}{lrrrl}
\toprule
Model & Phase~1 $d$ & Phase~2 $d$ & Ph2/Ph1 & Reading \\
\midrule
Phi-3.5-mini  & 6.78 & 3.97 & 0.59 & Lower, still large \\
Qwen2.5-0.5B  & 1.58 & 0.36 & 0.23 & Collapsed \\
\bottomrule
\end{tabular}
\end{table}

\section{Discussion}
\label{sec:discussion}

\noindent\textbf{How our size result fits earlier work.}
Our within-family trend is steep, while HarmBench saw no size effect inside families~\cite{mazeika2024harmbench} and one DeepSeek study saw the reverse for GCG~\cite{wu2025deepseekgpt}. We see four possible reasons. Our attack is a set of fixed templates, not an optimised attack. Our small models are much smaller (0.5B--3B) than most HarmBench models (7B and up). Our Llama family mixes two generations. And decoding differs between our local and hosted models. Our results show that size matters a lot for template attacks on small models; they do not show that it matters for every attack.

\noindent\textbf{Reporting checklist for safety verification.}
Our results suggest six practices.
\begin{enumerate}
\item Report absolute Phase~2 ASR with confidence intervals next to the Phase~1 baseline, and the log-odds shift if models of different strength are compared.
\item Report the judge's error rates per model, from a human-checked sample, and show how conclusions change when they are corrected.
\item Report RCF (or another measure of incoherent replies) next to ASR, and report raw counts and valid-response denominators.
\item Use the same decoding rules for every model (for example greedy decoding and 512 new tokens), pin the provider for gateway-served models, repeat hosted runs at least three times, and record model names and query dates, since hosted models drift and are retired.
\item Do not use parameter count as a stand-in for safety; weigh alignment reports and training history.
\item Record evaluator location and confirm it with an IP lookup, but do not expect it to matter at the 5--8 point level, and remember that a gateway may hide it from the provider.
\end{enumerate}

\noindent\textbf{What the geolocation null means.}
We find no detectable effect of the exit city on ASR at our resolution. Earlier work found location effects for moderation of sensitive topics~\cite{lipphardt2026basingse}, so a null for English jailbreak prompts is a different and narrower claim. Because most models were reached through a gateway, the result mainly shows that the exit city does not change what the gateway delivers. It does not rule out effects that a direct provider connection would reveal, effects below the detection limits, other cities, other topics, or non-English prompts.

\section{Limitations}
\label{sec:limits}
\begin{itemize}
\item \textbf{Resolution.} 400 prompts per phase give intervals up to $\pm$5 points; per-category numbers (about 80 prompts) are up to $\pm$11 points.
\item \textbf{Mixed-up factors.} Size is mixed up with alignment age, quantization, decoding rules, and how a model is served (local 4-bit vs.\ hosted). Hosted models may also have provider-side filters. Llama mixes 3.2 and 3.1, and Mixtral is the oldest model.
\item \textbf{Judging.} One classifier and one human reviewer, with $\sim$80 re-checked replies per model. Judge error differs by model, and the large models' Phase~1 ASRs are near the range where false positives matter.
\item \textbf{Denominators.} Overall ASR values are means of whole-percent category rates (up to about 0.5 points of rounding), and the RCF intervals do not match a fixed 400 responses (Appendix~\ref{app:setup}).
\item \textbf{Geolocation.} Five cities, three runs each, VPN exit nodes, gateway-served models, two Gemini cities missing, run-level analysis, and low power below the limits in Section~\ref{sec:geo}.
\item \textbf{Probe.} Two models, two components, last-token activations only, a quick budget; directional evidence, not a full interpretability result.
\item \textbf{Statistics.} The log-odds analysis treats samples as independent and does not model differences between models within a tier.
\item \textbf{Availability.} Model versions were retired or removed during or just before the study period (Gemini~1.5, Gemini~2.0 Flash, the OpenRouter Mixtral listing), so exact reruns may be impossible.
\end{itemize}

\section*{Ethics and Responsible Release}
We use only public prompts and templates and write no new attacks. We do not release harmful model outputs; we release aggregate counts, judge labels and activation arrays. Studying where small models fail helps defenders and deployers, but the same tables could guide an attacker to weaker models; we think the benefit to people who must choose models outweighs this.

\section{Conclusion}
\label{sec:conclusion}

Jailbreak vulnerability falls steeply with size inside the Llama and Qwen families, and wrapping shifts every model by roughly two to three units (1.7 to 2.9) on the log-odds scale, so plain testing understates absolute risk everywhere. Phi-3.5-mini and Mixtral show that parameter count is an unreliable proxy, although Phi-3.5-mini's lead is the least stable of our results. Judge agreement falls as ASR rises, so ASR gaps across sizes should be read with their measurement error in mind; the size gap itself survives our stress test. We detect no effect of the exit city on ASR, but our test is weak, and most models were reached through a gateway. For verification practice we recommend two-phase testing for small models, absolute wrapped ASR with intervals, judge error rates per model, identical decoding for all models, and the checklist above.

\section*{Acknowledgements}
This work was done at the Global South AI Safety Hackathon, June 2026, hosted by Apart Research. The author thanks the organisers for compute access. No outside funding was received, and the author has no competing interests.

\bibliographystyle{unsrt}
\bibliography{references}

\appendix
\newpage
\section*{Appendix}
\section{Statistical Details and Sensitivity Analyses}
\label{app:stats}

\noindent\textbf{Log-odds shift.}
For a model with Phase~1 and Phase~2 ASRs $\hat p_1$ and $\hat p_2$ on $n=400$ prompts each,
\[
s=\logit(\hat p_2)-\logit(\hat p_1),\qquad
\widehat{\mathrm{Var}}(s)=\frac{1}{n\hat p_1(1-\hat p_1)}+\frac{1}{n\hat p_2(1-\hat p_2)},
\]
where $\logit(p)=\ln\frac{p}{1-p}$ and $e^{s}$ is the factor by which the odds of success are multiplied. The variance treats the phases as independent; because the prompts are paired this is conservative or optimistic depending on their correlation, which we cannot compute without prompt-level data. Table~\ref{tab:stats} gives $s$, its standard error and $e^{s}$. The mean shift is 2.33 for the six small models and 2.00 for the six large ones. The difference is 0.33 with standard error 0.17, giving an approximate 95\% interval of 0.0 to 0.7. Inside the small tier the observed variance of the shifts (0.103) is above the average sampling variance (0.040), driven by Qwen2.5-0.5B and Phi-3.5-mini; inside the large tier it is below (0.054 vs.\ 0.133).

\begin{table}[h]
\caption{Log-odds shift from Phase~1 to Phase~2, approximate standard error, and odds multiple. Computed from Table~\ref{tab:results} with $n=400$.}
\label{tab:stats}
\centering
\small
\setlength{\tabcolsep}{4pt}
\begin{tabular}{l rrrrrr | rrrrrr}
\toprule
Metric & 
\rotatebox{90}{Llama-3.2-1B} & 
\rotatebox{90}{Llama-3.2-3B} & 
\rotatebox{90}{Qwen2.5-0.5B} & 
\rotatebox{90}{Qwen2.5-7B} & 
\rotatebox{90}{Gemma-2-2B} & 
\rotatebox{90}{Phi-3.5-mini} & 
\rotatebox{90}{Llama-3.1-70B} & 
\rotatebox{90}{Qwen2.5-72B} & 
\rotatebox{90}{Mixtral-8x7B} & 
\rotatebox{90}{DeepSeek-V3} & 
\rotatebox{90}{\gflash} & 
\rotatebox{90}{\gpro} \\
\midrule
$s$     & 2.30 & 2.28 & 2.90 & 2.39 & 2.23 & 1.91 & 1.79 & 2.06 & 2.39 & 2.01 & 2.03 & 1.73 \\
SE      & 0.17 & 0.19 & 0.19 & 0.20 & 0.17 & 0.26 & 0.31 & 0.41 & 0.23 & 0.41 & 0.39 & 0.41 \\
$e^{s}$ & 10.0 & 9.8  & 18.1 & 10.9 & 9.3  & 6.8  & 6.0  & 7.9  & 10.9 & 7.4  & 7.6  & 5.7  \\
\bottomrule
\end{tabular}
\end{table}

\noindent\textbf{Judge false positives.}
If the judge has sensitivity $\mathrm{se}$ and false-positive rate $\mathrm{fp}$, the observed ASR is $\hat p_{\text{obs}}=\mathrm{se}\,p+\mathrm{fp}\,(1-p)$, so the corrected ASR is $\hat p=(\hat p_{\text{obs}}-\mathrm{fp})/(\mathrm{se}-\mathrm{fp})$ (the Rogan--Gladen estimator). We do not know the judge's rates on our replies, so Table~\ref{tab:judgesens} uses $\mathrm{se}=0.90$ and $\mathrm{fp}=0.01$ as an assumption. The published false-positive rate of the JailbreakBench judge on its own validation set is about 12\%, but that set is not ours, and an easy refusal is much less likely to be misjudged than a borderline reply. Under this assumption the tier gap falls from 0.33 to 0.03, and if $\mathrm{fp}$ exceeded about 1.8\% the lowest Phase~1 ASRs would be explained entirely by false positives.

\begin{table}[h]
\caption{Log-odds shift $s$ as reported and after correcting for an assumed judge sensitivity of 0.90 and false-positive rate of 1\%. The rates are assumed, not measured.}
\label{tab:judgesens}
\centering
\small
\begin{tabular}{lrr}
\toprule
Model & $s$ as reported & $s$ corrected \\
\midrule
Llama-3.2-1B  & 2.30 & 2.57 \\
Llama-3.2-3B  & 2.28 & 2.49 \\
Qwen2.5-0.5B  & 2.90 & 3.27 \\
Qwen2.5-7B    & 2.39 & 2.60 \\
Gemma-2-2B    & 2.23 & 2.48 \\
Phi-3.5-mini  & 1.91 & 2.14 \\
\midrule
Llama-3.1-70B & 1.79 & 2.10 \\
Qwen2.5-72B   & 2.06 & 2.80 \\
Mixtral-8x7B  & 2.39 & 2.62 \\
DeepSeek-V3   & 2.01 & 2.74 \\
\gflash       & 2.03 & 2.66 \\
\gpro         & 1.73 & 2.44 \\
\midrule
Small-tier mean & 2.33 & 2.59 \\
Large-tier mean & 2.00 & 2.56 \\
Gap             & 0.33 & 0.03 \\
\bottomrule
\end{tabular}
\end{table}

\noindent\textbf{Stress test for incoherent replies.}
Table~\ref{tab:stress} subtracts each model's RCF from its Phase~2 ASR. This is a stress test and not a bound: judge errors may also occur among coherent replies, and the overlap between RCF and ASR is unknown (it is at least 1.9 points for Qwen2.5-0.5B, whose ASR and RCF sum to more than 100\%).

\begin{table}[h]
\caption{Stress test: Phase~2 ASR with each model's RCF subtracted.}
\label{tab:stress}
\centering
\small
\begin{tabular}{lrrr}
\toprule
Model & Ph2 ASR & RCF & ASR $-$ RCF \\
\midrule
Llama-3.2-1B  & 65.6 & 24.7 & 40.9 \\
Llama-3.2-3B  & 53.8 & 17.4 & 36.4 \\
Qwen2.5-0.5B  & 70.8 & 31.1 & 39.7 \\
Qwen2.5-7B    & 51.2 & 12.3 & 38.9 \\
Gemma-2-2B    & 65.2 & 21.6 & 43.6 \\
Phi-3.5-mini  & 25.4 &  9.2 & 16.2 \\
\midrule
Llama-3.1-70B & 17.4 &  2.1 & 15.3 \\
Qwen2.5-72B   & 12.6 &  1.4 & 11.2 \\
Mixtral-8x7B  & 41.0 &  5.8 & 35.2 \\
DeepSeek-V3   & 12.0 &  0.8 & 11.2 \\
\gflash       & 13.4 &  1.0 & 12.4 \\
\gpro         &  9.4 &  0.6 &  8.8 \\
\bottomrule
\end{tabular}
\end{table}

\noindent\textbf{Detection limits for the geolocation test.}
Let $\sigma$ be the binomial standard error of one city--model cell with $n=400$ responses: 2.3, 2.5 and 1.6 points for Phi-3.5-mini, Qwen2.5-7B and DeepSeek-V3. (i)~For one city against the mean of the other four, the standard error of the difference is $\sigma\sqrt{1+1/4}$, so the smallest shift detected with 80\% power at the 5\% level is $(1.96+0.84)\,\sigma\sqrt{1.25}$: 7.2, 7.8 and 5.1 points. (ii)~For the omnibus test on the standard deviation of five city means, the 95\% critical value is $\sigma\sqrt{\chi^2_{4,0.95}/4}=1.54\,\sigma$ under the null. If city effects have standard deviation $\tau$, the statistic is $\sqrt{\sigma^2+\tau^2}\sqrt{\chi^2_4/4}$, so power of 50\% needs $\tau\approx1.35\,\sigma$ and power of 80\% needs $\tau\approx2.18\,\sigma$: 3.1 to 5.0 points for Phi-3.5-mini, 3.4 to 5.4 for Qwen2.5-7B, and 2.2 to 3.5 for DeepSeek-V3. These limits assume independent responses; a paired design would lower them.

\section{Setup Details}
\label{app:setup}

\begin{table}[h]
\caption{Category mapping. HarmBench labels are from its semantic categories; fraud has no HarmBench label.}
\label{tab:catmap}
\centering
\small
\begin{tabular}{ll}
\toprule
Study category & Source label \\
\midrule
Cybercrime & HarmBench: cybercrime\_intrusion \\
Misinformation & HarmBench: misinformation\_disinformation \\
Harassment & HarmBench: harassment\_bullying \\
CBRN-adjacent & HarmBench: chemical\_biological \\
Fraud & none in HarmBench (a subtype of ``illegal''); AdvBench by keyword matching \\
Left out & HarmBench: copyright (needs hash checks), general\_harm \\
\bottomrule
\end{tabular}
\end{table}

\noindent\textbf{Denominators.}
The reported RCF percentages are not all multiples of 0.25, so they are not all counts out of exactly 400 (for example, 24.7\% would be 98.8 of 400). The overall ASR values in Table~\ref{tab:results} are means of five whole-percent category values (Figure~\ref{fig:fullheat}), so each is a multiple of 0.2 (for example, 65.6\% would be 262.4 of 400) and can differ from the true count by up to about 0.5 points. For the RCF intervals, two explanations fit: denominators below 400 after empty or failed generations were dropped, or values that are means over repeated runs. The reported Wilson intervals favour the first. Back-solving the denominator from the interval gives about 370 for Llama-3.2-1B (RCF 24.7\%) and 375 for Qwen2.5-0.5B (RCF 31.1\%), but about 430 to 600 for the lowest-RCF rows such as Llama-3.1-70B and DeepSeek-V3, so the intervals were not computed with one common denominator. Table~\ref{tab:counts} shows the nearest counts out of 400 as a reference, together with Wilson 95\% intervals for Phase~1 and Phase~2 ASR computed at $n=400$.

\begin{table}[h]
\caption{Nearest counts out of 400 for the reported percentages, and Wilson 95\% intervals (in percent, $n=400$) for Phase~1 and Phase~2 ASR. Because the ASR percentages are means of whole-number category values, ASR counts are approximate to about $\pm$2 responses and the intervals are approximate.}
\label{tab:counts}
\centering
\small
\begin{tabular}{lrrrrr}
\toprule
Model & Ph1 count & Ph1 95\% CI & Ph2 count & Ph2 95\% CI & RCF count \\
\midrule
Llama-3.2-1B  & 64  & 12.7--19.9 & 262 & 60.8--70.1 & 99 \\
Llama-3.2-3B  & 42  & 7.9--14.0  & 215 & 48.9--58.6 & 70 \\
Qwen2.5-0.5B  & 47  & 9.0--15.3  & 283 & 66.2--75.0 & 124 \\
Qwen2.5-7B    & 35  & 6.4--12.0  & 205 & 46.3--56.1 & 49 \\
Gemma-2-2B    & 67  & 13.5--20.8 & 261 & 60.4--69.7 & 86 \\
Phi-3.5-mini  & 19  & 3.1--7.4   & 102 & 21.4--29.9 & 37 \\
\midrule
Llama-3.1-70B & 14  & 2.0--5.7   & 70  & 14.0--21.4 & 8 \\
Qwen2.5-72B   & 7   & 0.9--3.6   & 50  & 9.7--16.2  & 6 \\
Mixtral-8x7B  & 24  & 4.1--8.8   & 164 & 36.3--45.9 & 23 \\
DeepSeek-V3   & 7   & 0.9--3.6   & 48  & 9.2--15.6  & 3 \\
\gflash       & 8   & 1.0--3.9   & 54  & 10.4--17.1 & 4 \\
\gpro         & 7   & 0.9--3.6   & 38  & 6.9--12.7  & 2 \\
\bottomrule
\end{tabular}
\end{table}

\noindent\textbf{Request budget.}
The hosted part of the study needs about 12{,}000 requests: 6 hosted models $\times$ 800 prompts, plus 18 city--model cells $\times$ about 400 responses. OpenRouter's free models allow 20 requests per minute and 50 per day, or 1{,}000 per day after at least \$10 of credits has ever been bought\footnote{\url{https://openrouter.ai/docs/api_reference/limits}}, so a study of this size on free models alone would take between 12 and 240 days. We therefore describe the infrastructure as low-cost: Mixtral-8x7B and, for the location test, Phi-3.5-mini are paid models on OpenRouter. Gemini's free-tier limits were cut in December 2025 and are no longer published per model, so we do not estimate that part.

\section{Extended Scale Analysis}
\label{app:scale}

\noindent\textbf{Uplift against size.}
Figure~\ref{fig:uplift} plots $\Delta$ against log parameter count, with bubble area showing RCF. Among small models, $\Delta$ and RCF rise together. Both follow Phase~2 ASR closely, so this is not an independent effect, and we have not tested any shared cause. Phi-3.5-mini lies below the small-model trend in both uplift and RCF, and Mixtral lies above the large-model trend.

\begin{figure}[h]
\centering
\begin{subfigure}[t]{0.62\linewidth}
  \includegraphics[width=\linewidth]{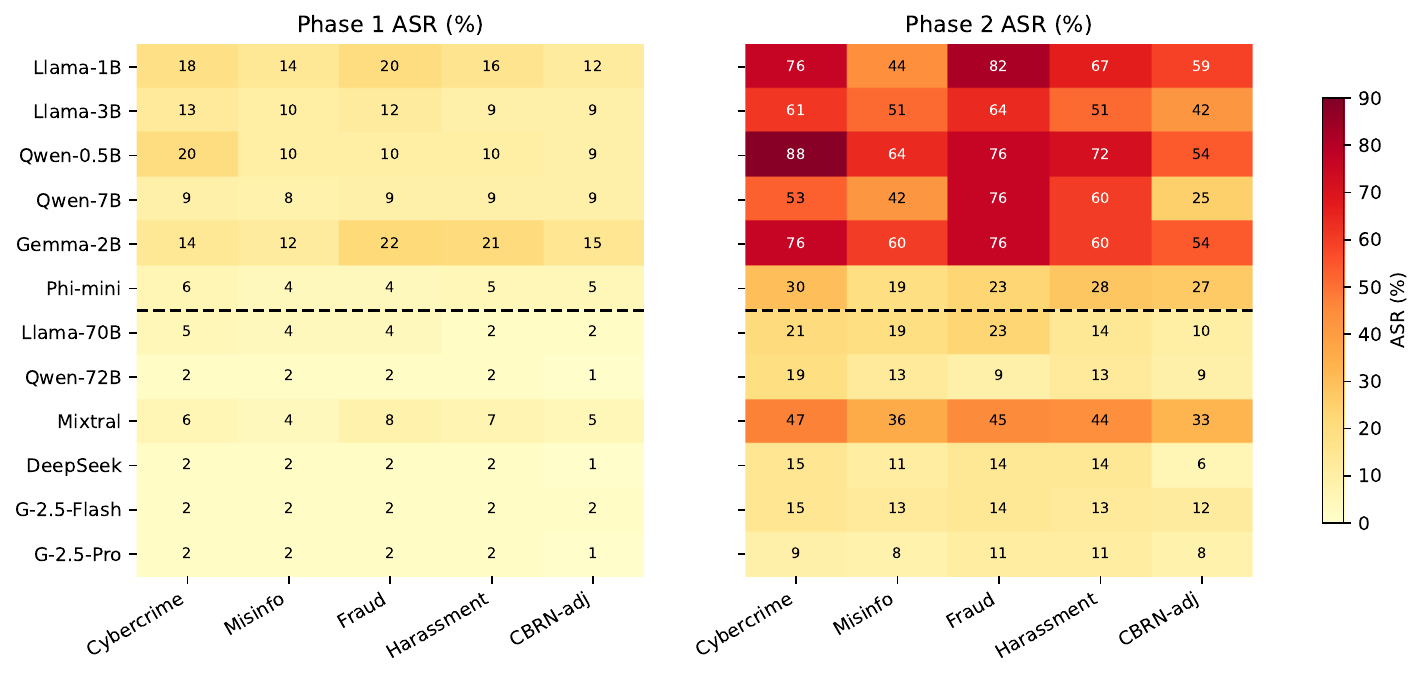}
  \caption{Phase~1 (left) and Phase~2 (right) ASR for all twelve models by harm category. The dashed line separates small (top) and large (bottom) tiers. The colour scale is shared.}
  \label{fig:fullheat}
\end{subfigure}
\hfill
\begin{subfigure}[t]{0.36\linewidth}
  \includegraphics[width=\linewidth]{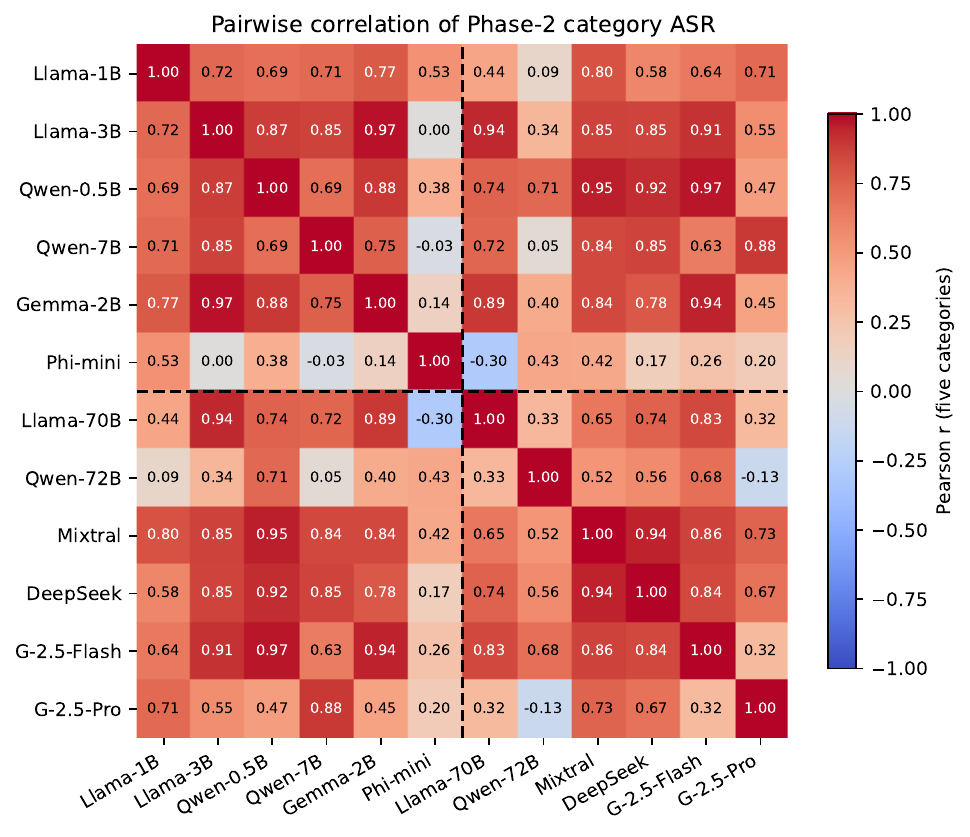}
  \caption{Pairwise Pearson $r$ of Phase~2 ASR across five harm categories for all twelve models. Dashed lines mark the small/large boundary.}
  \label{fig:corr}
\end{subfigure}
\caption{Full ASR heatmap (left) and pairwise failure correlation (right).}
\label{fig:heatcorr}
\end{figure}

\noindent\textbf{Full ASR heatmap.}
Figure~\ref{fig:fullheat} shows all twelve models. In Phase~2 the cybercrime column is darkest for Qwen2.5-0.5B and Phi-3.5-mini. Fraud is darkest for Llama-3.2-1B, Llama-3.2-3B and Qwen2.5-7B, and Gemma-2-2B ties the two. CBRN-adjacent is the lightest column for most models, but not for Llama-3.2-1B or Phi-3.5-mini, where misinformation is. Gemma-2-2B has high Phase~1 fraud and harassment ASR (22\% and 21\%) compared with Llama-3.2-3B (12\% and 9\%); with 80 prompts per cell this is suggestive only.

\noindent\textbf{Uplift by harm category.}
Figure~\ref{fig:deltacat} splits the uplift by category. Cybercrime or fraud gives the largest uplift for every small model (fraud for Llama-3.2-1B, Llama-3.2-3B and Qwen2.5-7B; cybercrime for the others). Together they make up 42--52\% of the summed uplift in small models, against 40\% if all five categories were equal, so the excess is modest. Phi-3.5-mini has the most even profile among small models.

\section{Composite Safety Score (descriptive only)}
\label{app:score}

For readers who want one number, we combine four sub-scores with a geometric mean, which punishes a model that is poor on any one: (i)~Phase~2 ASR, (ii)~RCF, (iii)~uplift $\Delta$, and (iv)~the cybercrime$-$CBRN-adjacent gap. Each sub-score is min--max scaled across the twelve models to 0--100 (100 = safest or most reliable). Because the scaling is relative to this pool, the best model on each sub-score gets 100 by construction, and any sub-score of 0 forces a composite of 0. The sub-scores are strongly related ($\Delta$ and Phase~2 ASR correlate at 0.996), so the composite counts ASR twice. We use it only to show each model's profile, not as a calibrated metric.

\begin{figure}[h]
\centering
\begin{subfigure}[t]{0.49\linewidth}
  \includegraphics[width=\linewidth]{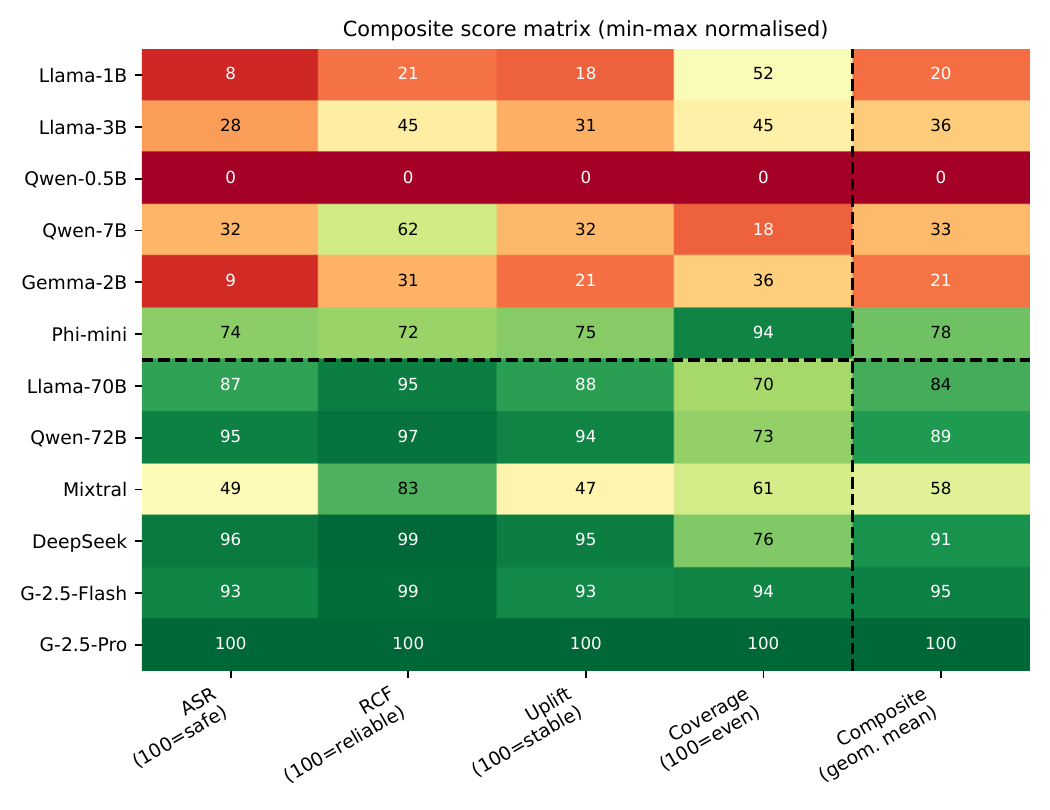}
  \caption{Composite score matrix (sub-scores min--max scaled to 0--100; the right column is the geometric mean). The dashed horizontal line separates small and large tiers.}
  \label{fig:score_matrix}
\end{subfigure}
\hfill
\begin{subfigure}[t]{0.49\linewidth}
  \includegraphics[width=\linewidth]{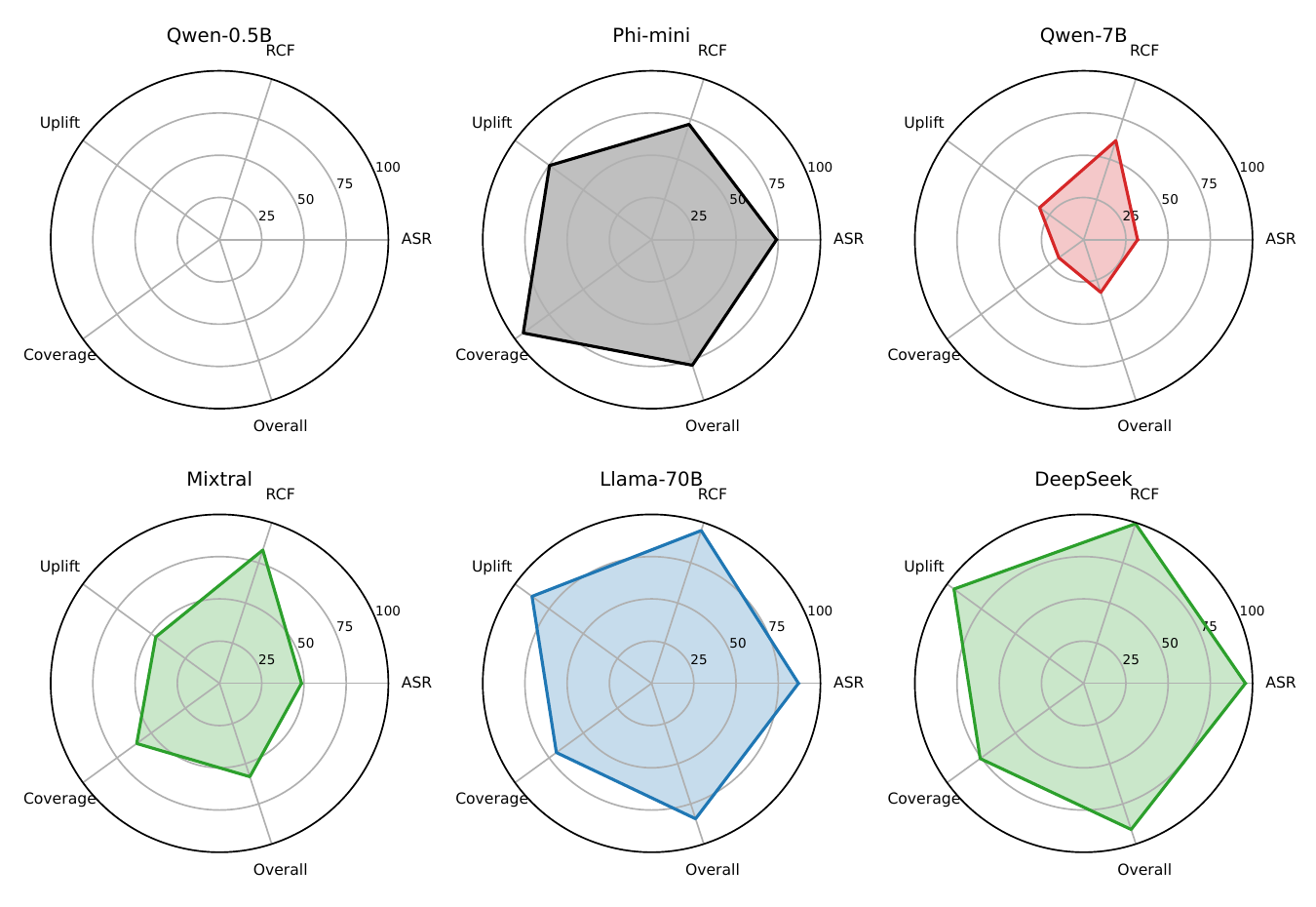}
  \caption{Radar charts for six models (0--100, higher is safer).}
  \label{fig:radar}
\end{subfigure}
\caption{Composite safety score: matrix (left) and radar view (right).}
\label{fig:score}
\end{figure}

Figure~\ref{fig:score_matrix} shows the matrix and Figure~\ref{fig:radar} the radar view. \gpro{} scores 100, \gflash{} about 95 and DeepSeek-V3 about 91. Among small models Phi-3.5-mini scores highest (about 78), closer to the large tier than to its peers, while Qwen2.5-0.5B scores 0 and Llama-3.2-1B about 20. Mixtral scores about 58, well below the two large dense models (84 and 89). Coverage is the weakest sub-score for the large dense models (70 and 73).

\section{Pairwise Model Failure Correlation}
\label{app:corr}

Figure~\ref{fig:corr} shows Pearson correlations between the Phase~2 ASR vectors of each pair of models across the five harm categories. With only five points per vector these correlations are very noisy (an $r$ of 0.9 has a 95\% interval of roughly 0.1 to 0.99), so we treat them as exploratory. Leaving out Phi-3.5-mini, small-model pairs correlate between 0.69 and 0.97. Strong links also cross tiers: Qwen2.5-0.5B with \gflash{} (0.97), Mixtral (0.95) and DeepSeek-V3 (0.92). The large tier is not a tight block: Llama-3.1-70B and Qwen2.5-72B correlate at 0.33, and Qwen2.5-72B and \gpro{} at $-0.13$. Phi-3.5-mini is the least correlated model with all others ($-0.30$ to $0.53$); its profile is nearly flat (19--30\%), so small changes dominate $r$. The data support a shared category pattern for most models, with Phi-3.5-mini and Qwen2.5-72B as exceptions. They do not support a tier-based failure profile.

\section{RCF by Harm Category}
\label{app:rcfcat}

Table~\ref{tab:rcfcat} and Figure~\ref{fig:rcfcat} give Phase~2 RCF by category. Cybercrime has the highest RCF for all six small models. The spread inside a model is small (about 6 points for Qwen2.5-0.5B and Llama-3.2-1B), below the resolution at 80 prompts per category, so only the consistent direction across models is informative. Cybercrime is also among the highest-ASR categories (highest or second highest for five of six small models), so confusion and compliance go up together across categories. RCF should still be reported per category: a model with 31\% average RCF can hide 35.0\% on cybercrime next to 28.9\% on misinformation (Qwen2.5-0.5B).

\begin{table}[h]
\caption{Phase~2 RCF (\%) by model and harm category.}
\label{tab:rcfcat}
\centering
\small
\begin{tabular}{lrrrrr}
\toprule
Model & Cybercrime & Misinfo & Fraud & Harassment & CBRN-adj \\
\midrule
Llama-3.2-1B  & 28.1 & 22.3 & 26.0 & 25.5 & 21.5 \\
Llama-3.2-3B  & 19.1 & 15.7 & 18.1 & 17.0 & 17.1 \\
Qwen2.5-0.5B  & 35.0 & 28.9 & 31.7 & 30.2 & 29.7 \\
Qwen2.5-7B    & 13.5 & 10.8 & 12.4 & 12.0 & 12.8 \\
Gemma-2-2B    & 24.3 & 19.5 & 22.0 & 21.2 & 20.9 \\
Phi-3.5-mini  & 10.1 &  8.0 &  9.3 &  9.1 &  9.3 \\
\midrule
Llama-3.1-70B &  2.2 &  1.9 &  2.3 &  2.0 &  2.1 \\
Qwen2.5-72B   &  1.5 &  1.2 &  1.4 &  1.3 &  1.6 \\
Mixtral-8x7B  &  6.2 &  5.1 &  6.0 &  5.8 &  5.9 \\
DeepSeek-V3   &  0.9 &  0.7 &  0.8 &  0.7 &  0.9 \\
\gflash       &  1.1 &  0.9 &  1.0 &  1.0 &  1.0 \\
\gpro         &  0.6 &  0.5 &  0.6 &  0.6 &  0.7 \\
\bottomrule
\end{tabular}
\end{table}

\section{Extended Geolocation Analysis}
\label{app:geoext}

\begin{figure}[h]
\centering
\begin{subfigure}[t]{\linewidth}
  \centering
  \includegraphics[width=0.95\linewidth]{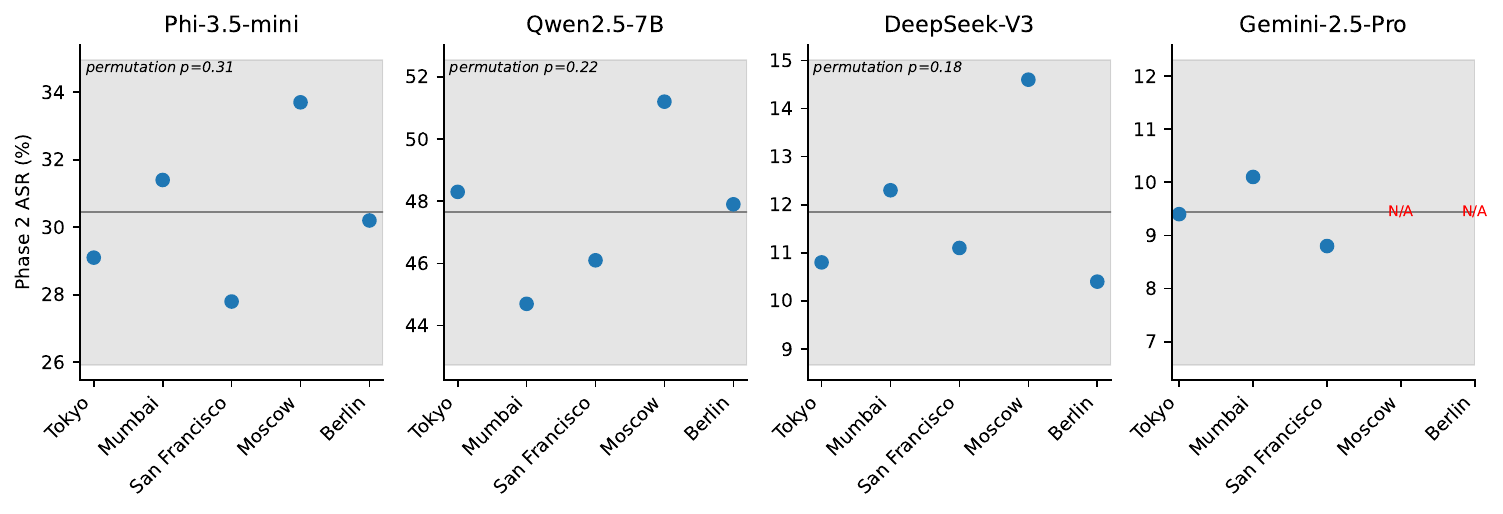}
  \caption{Phase~2 ASR by VPN exit city. Points are the city means from Table~\ref{tab:geo}; the shaded band is $\pm1.96$ binomial standard errors for 400 responses around each model's mean.}
  \label{fig:geostrip}
\end{subfigure}

\vspace{4pt}
\begin{subfigure}[t]{\linewidth}
  \centering
  \includegraphics[width=0.95\linewidth]{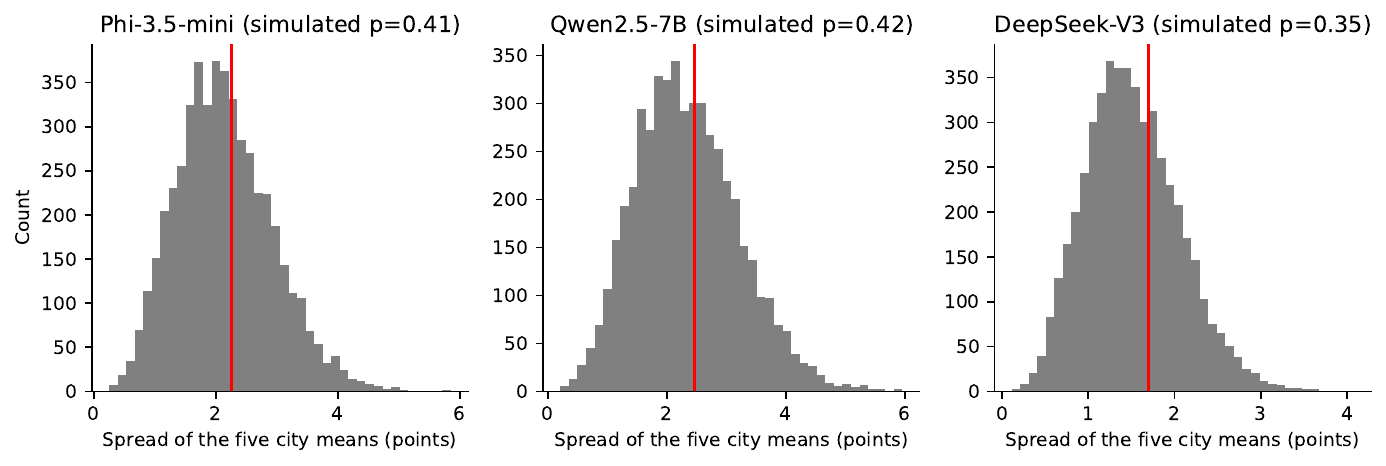}
  \caption{Simulated null distributions of the spread of city-mean ASR (5000 draws, 400 independent responses per cell). Red line: observed spread.}
  \label{fig:permnull}
\end{subfigure}
\caption{Geolocation: city means against binomial noise (top) and simulated null distributions (bottom).}
\label{fig:geo}
\end{figure}

\noindent\textbf{City means against the expected noise.}
Figure~\ref{fig:geostrip} shows Phase~2 ASR by exit city. The shaded band is the 95\% range expected from binomial noise alone for 400 independent responses around each model's mean ASR. Every observed city mean lies inside its band, for all three fully testable models and for the three usable \gpro{} cities. The \gpro{} panel shows N/A where no data could be collected.

\noindent\textbf{Null distributions.}
Figure~\ref{fig:permnull} shows the distribution of the spread of the five city means when each cell is simulated as 400 independent responses at the model's mean ASR (5000 draws). The observed spread is larger than about 59\%, 58\% and 65\% of the simulated values for Phi-3.5-mini, Qwen2.5-7B and DeepSeek-V3, which gives $p\approx0.41$, $0.42$ and $0.35$. The 95th percentile of the simulated spread is about 3.5, 3.8 and 2.5 points. The run-level permutation $p$-values reported in Table~\ref{tab:geo} (0.31, 0.22 and 0.18) are smaller than these, and both sets of values are at least 0.18. The detection limits that follow from this noise are derived in Appendix~\ref{app:stats}; they are a limit of the design on low-cost infrastructure, not a threshold fixed in advance.

\section{Contrastive Probe Details}
\label{app:probe}

Activations come from the final transformer layer. PCA is fitted separately on the pooled refusal and compliance activations of each (model, phase) condition, so the axes favour neither class but are not comparable across panels. A linear discriminant (scikit-learn \texttt{LinearDiscriminantAnalysis}, SVD solver, no shrinkage) is fitted on the top-two projection and checked by leave-one-out cross-validation ($n=240$). Cohen's $d$ uses the pooled within-class standard deviation of the discriminant scores. All probe runs use the same Colab T4 GPU as local inference. The released code also draws the two-component scatter plots of the activations.

The probe is directional evidence, not a causal test. It shows that the \emph{linear} refusal/compliance split is more disrupted in Phase~2 for Qwen2.5-0.5B than for Phi-3.5-mini, in line with the behaviour results. It does not show that this is the main reason jailbreaks work, and activation patching would be needed for a causal claim. Because safety uses more than one direction~\cite{pan2025hidden}, a two-component probe may miss part of the picture.

\section{Classifier Calibration}
\label{app:classifier}

\begin{figure}[h]
\centering
\includegraphics[width=0.95\linewidth]{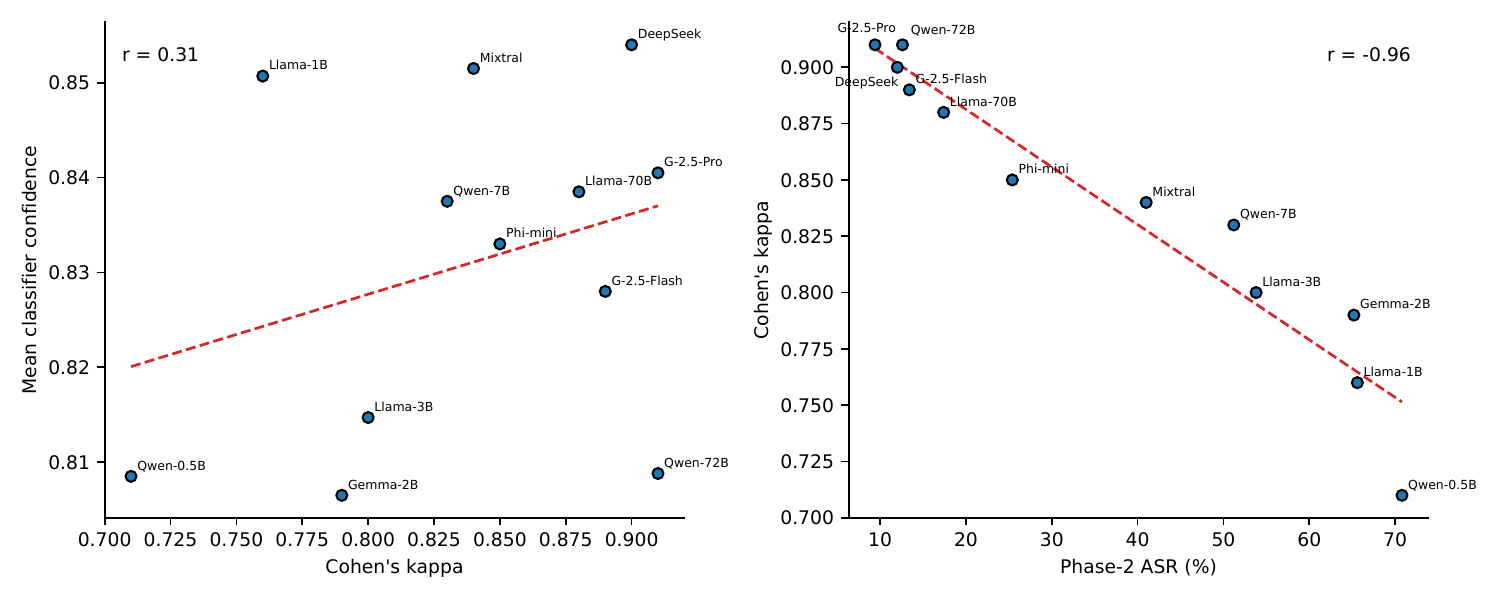}
\caption{\textit{Left:} classifier confidence against $\kappa$ ($r=0.31$). \textit{Right:} Phase~2 ASR against $\kappa$ ($r=-0.96$). Dashed lines are OLS fits.}
\label{fig:calibration}
\end{figure}

The JailbreakBench-style classifier is not perfect. We re-checked 10\% of replies and computed Cohen's $\kappa$ for each model (Figure~\ref{fig:calibration}). $\kappa$ is only weakly related to the classifier's average confidence ($r=0.31$, $n=12$, not significant), so confidence is a poor guide to human agreement. $\kappa$ falls strongly as Phase~2 ASR rises ($r=-0.96$): the high-ASR small models, which also have the highest RCF, produce the most unclear replies. This link is mixed up with model size and serving mode, and each $\kappa$ rests on about 80 replies, so differences of a few hundredths mean little. Qwen2.5-0.5B has the lowest $\kappa$ (0.71), probably because 31.1\% of its replies are incoherent.

Because confidence does not track agreement, automated judges are least reliable on the most vulnerable models, and their confidence gives no warning. This agrees with recent audits of safety judges~\cite{schwinn2026coinflip,gao2026judge,eiras2025knowthyjudge}. We recommend human spot-checks of at least 10\% of replies, sampled by model and by classifier confidence, for any small-model safety test.

\end{document}